\documentclass[a4paper]{jpconf}
\usepackage{graphicx}

\begin{document}
\title{Measurement of beam energy dependent nuclear modification factors at STAR}

\author{Stephen P. Horvat (for the STAR Collaboration)}

\address{Department of Physics, 217 Prospect Street, New Haven, CT 06511-8499}
\ead{stephen.horvat@yale.edu}

\begin{abstract}
The nuclear modification factors $R_{ \mathrm{AA}}$ and $R_{ \mathrm{CP}}$ have been used to measure medium-induced suppression in heavy-ion collisions at $\sqrt{s_{_{ \mathrm{NN}}}}$ = 200 GeV which was among the earliest evidence for the existence of a strongly interacting medium called a quark-gluon plasma (QGP). Nuclear modification factors for asymmetric collisions ($R_{ \mathrm{dA}}$) have measured the Cronin Effect, an enhancement of high transverse momentum particle yields in deuteron-gold collisions relative to proton-proton collisions. A similar enhancement is observed in data presented in these proceedings and competes with the quenching caused by partonic energy loss in the QGP. In these proceedings we will present charged-hadron $R_{ \mathrm{CP}}$ at mid-rapidity for $\sqrt{s_{_{ \mathrm{NN}}}}$ = 7.7 - 62.4 GeV as well as identified $\pi^{+}$, $K^{+}$, and $p$ $R_{ \mathrm{CP}}$.  Comparisons to HIJING motivate possible methods for disentangling competing modifications to nuclear transverse momentum spectra.
\end{abstract}

\section{Introduction}
The RHIC beam energy scan (BES) is a program to collide Au+Au ions at various collision energies  in order to explore the QCD phase diagram; searching for a possible critical point and for a  phase boundary marked by the disappearance of key signatures for the formation of a QGP [1].  The nuclear modification factor provides one of these signatures.  The ratio of transverse momentum ($p_{ \mathrm{T}}$) differentiated spectra from central over peripheral collisions and scaled by the mean number of binary $p$+$p$-like collisions in each event is called the nuclear modification factor and is denoted by $R_{ \mathrm{CP}}$.  If $p$+$p$ collisions are used for the reference instead then the nuclear modification factor is denoted by $R_{ \mathrm{AA}}$.  In the presence of a QGP, high-$p_{ \mathrm{T}}$ particles are quenched, transferring energy to lower momentum particles, causing the nuclear modification factor to be less than unity at high $p_{ \mathrm{T}}$, or suppressed [2-6].   Quenching competes with any effects that would cause enhancement, such as radial boosts or the Cronin Effect [7].  The Cronin Effect has also been observed by STAR as the enhancement of the nuclear modification factor in asymmetric $d$+$Au$ collisions at $\sqrt{s_{_{ \mathrm{NN}}}}$ = 200 GeV [8].  These spectra can be influenced by the spectators, non-interacting nucleons in the collision system, so that a peripheral collision does not equate directly to a $p$+$p$ collision due to cold nuclear matter (CNM) effects.  The goal of this analysis is to determine at what beam energy suppression turns off, and to begin disentangling the causes and relative effects of quenching and enhancement.

\begin{figure}
\begin{center}
\includegraphics[width=39pc]{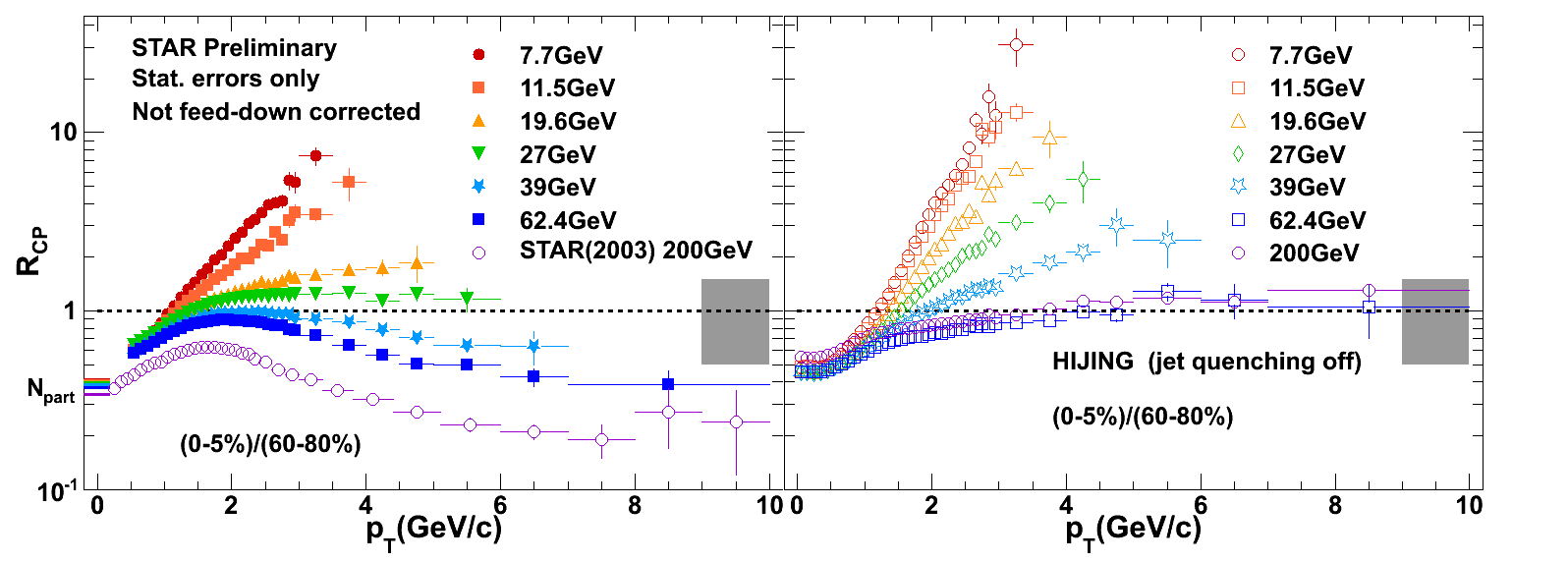}
\end{center}
\caption{\label{label}(Color online) Efficiency corrected charged hadron $R_{ \mathrm{CP}}$ (left) for RHIC BES energies.  $p_{ \mathrm{T}}$ dependent errors are statistical only.  The gray error band corresponds to the $p_{ \mathrm{T}}$ independent uncertainty in $N_{ \mathrm{coll}}$ scaling.  Charged $R_{ \mathrm{CP}}$ from HIJING simulated events (right) with jet quenching turned off.}
\end{figure}

\section{Charged-hadron $R_{ \mathrm{CP}}$}
$Au$+$Au$ collisions near the center of the detector were analyzed.  A particle track was used if its distance of closest approach ($dca$) was less than one centimeter in order to reduce feed-down contamination.  Trigger efficiency, tracking efficiency, and acceptance corrected charged-hadron spectra from mid-rapidity $Au$+$Au$ collisions at $\sqrt{s_{_{ \mathrm{NN}}}}$ =  7.7, 11.5, 19.6, 27, 39, and 62.4 GeV were produced for 0-5\% central and 60-80\% peripheral collisions in the STAR detector.  The spectra were scaled by binary collisions with the scale factors obtained from a Monte Carlo Glauber model [9].  Taking the ratio of these scaled spectra for each energy gives the $R_{ \mathrm{CP}}(\sqrt{s_{_{ \mathrm{NN}}}},p_{ \mathrm{T}})$ shown in Fig. 1 (left) along with STAR's published 200 GeV result [2].  These spectra were not feed-down corrected and were taken from -0.5 \textless\ $\eta$ \textless\ 0.5.  The global systematic uncertainty is dominated by the uncertainty in the centrality selection for the peripheral bins, which is used in the Glauber calculation and presents as an uncertainty on the binary collision scale factor.  The same methods that produced the lower beam energy results were used to produce a 200 GeV result from 2010 data.  This measurement disagreed with the feed-down corrected result from 2003 (Fig. 1 left) by 20\% and so this was folded into the overall systematic uncertainty of the other results (Fig. 1 gray box).  The cause of this discrepancy is under investigation.  The efficiency correction is based on single particle embedding in the 39 GeV data set for $\pi^{\pm}$, $K^{\pm}$, and $p^{\pm}$ separately which were than combined and weighted by their relative yields for a charged hadron efficiency.  The efficiency correction was extrapolated to the other data sets by making the assumption that the efficiency is the same for each particle at the same $p_{ \mathrm{T}}$ and from the same multiplicity bin.  This assumption was tested by producing the efficiencies from two data sets, 39 GeV from 2010 and 27 GeV from 2011, and ensuring that the predicted efficiency from the 39 GeV data set matched the 27 GeV efficiency.  Then differences in acceptance due to detector performance between beam energies were accounted for by using stable portions of the detector as a reference.  HIJING 1.35 [10] with jet quenching turned off was used to produce more than 10M collisions for each beam energy.  The motivation for running the simulator with jet quenching turned off was the expectation that at sufficiently low beam energies, where medium-induced jet quenching has minimal effect, there would be a quantitative agreement between the charged hadron $R_{ \mathrm{CP}}$ from simulation and data.  If the simulation and the data agreed at low beam energies but deviated at higher beam energies then the beam energies where the deviation occurred could be considered candidates for the beam energies where a QGP is formed.  Centrality selection was done using the same method as for the data; namely, counting the number of final state charged hadrons in -0.5 \textless\ $\eta$ \textless\ 0.5 and then determining the 0-5\% most central data as being the 5\% of events with the highest multiplicity, and so forth for the other centralities.  The result is shown in Fig. 1 on the right.  Again, $R_{ \mathrm{CP}}$ at lower beam energies is enhanced, although we do not see a quantitative agreement with charged hadron $R_{ \mathrm{CP}}$ from data.  We do not see suppression at higher collision energies, as expected since quenching was turned off.  By running HIJING with jet quenching on and off and comparing with AMPT and other models we hope to disentangle the relative contributions of jet quenching, CNM effects, and possible contributions from radial flow or final state scattering.  The results in Fig. 1 (left) are consistent with suppression for $\sqrt{s_{_{ \mathrm{NN}}}}$ $\geq$ 39 GeV.  This does not preclude medium induced energy losses at lower energies since other effects could be overwhelming this signature.  The advantage for using charged hadron $R_{ \mathrm{CP}}$ is that you can measure spectra to higher transverse momenta that you could not reach with particle identification.  It was also considered that plotting $R_{ \mathrm{CP}}$ vs. $x_{ \mathrm{T}}$ rather than $p_{ \mathrm{T}}$ might reveal trends in the data that were independent of collision energy.  $x_{ \mathrm{T}}$ is defined as $x_{ \mathrm{T}} = 2*p_{ \mathrm{T}}/\sqrt{s_{_{ \mathrm{NN}}}}$.  This sort of scaling was applied to spectra previously [8] where it revealed $\sqrt{s_{_{ \mathrm{NN}}}}$ independent trends at high $p_{ \mathrm{T}}$.  Such a scaling is shown in Fig. 2, using the data from Fig. 1,  and does not reveal any such trends.

\begin{figure}
\includegraphics[width=20pc]{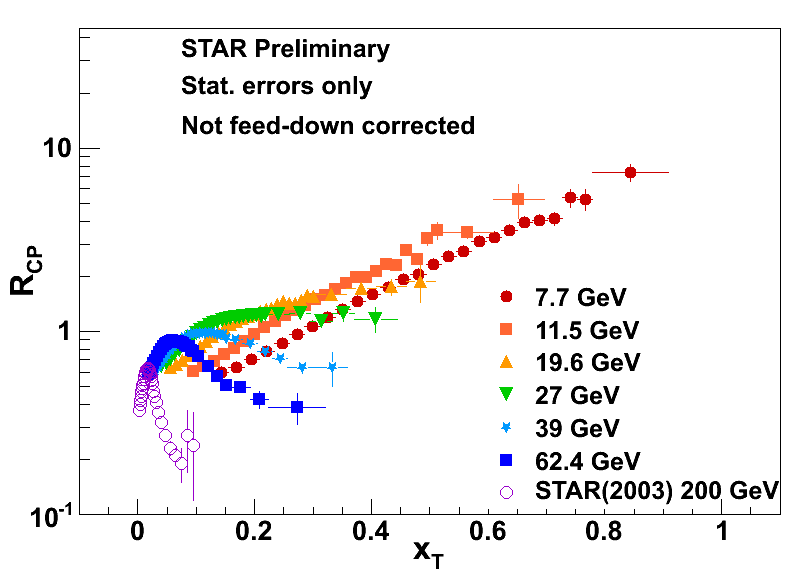}\hspace{2pc}%
\begin{minipage}[b]{15.3pc}\caption{\label{label}(Color online) The data from Fig. 1 (left) plotted vs. $x_{ \mathrm{T}}$ rather than $p_{ \mathrm{T}}$.}
\end{minipage}
\end{figure}

\begin{figure}
\begin{center}
\includegraphics[width=36pc]{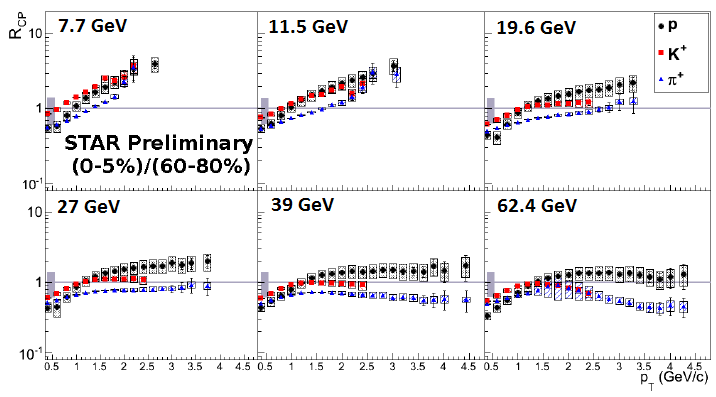}
\end{center}
\caption{\label{label}(Color online) $R_{ \mathrm{CP}}$ for identified $p$, $K^+$, and $\pi^+$ for RHIC BES energies.  The boxes are $p_{ \mathrm{T}}$ dependent systematic uncertainties due to particle identification while the $p_{ \mathrm{T}}$ independent uncertainty is from $N_{coll}$ scaling.}
\end{figure}

\section{Identified hadron $R_{ \mathrm{CP}}$} 
Particle yields were extracted from a simultaneous fit to $dE/dx$ distributions measured in the STAR Time Projection Chamber and time of flight distributions measured in the STAR Time of Flight detector for each centrality and $p_{ \mathrm{T}}$ bin at each beam energy.  The functions used to extrapolate fit parameters for particle identification were varied in order to obtain the systematic errors for the high $p_{ \mathrm{T}}$ bins.  Efficiency corrections were obtained through track embedding and have a 5\% systematic error associated with them.  The differences between the published 62.4 GeV results [11] and those presented here were taken as a point by point systematic error and applied to all beam energies.  The cause of this discrepancy is under investigation.  The result (Fig. 3) is qualitatively consistent with published results [11] in that pions are less enhanced than protons, suggesting that pions may serve as a better gauge for jet quenching within the $p_{ \mathrm{T}}$ range available through particle identification.  Considering 2.5 GeV/c \textless\ $p_{ \mathrm{T}}$ \textless\ 4 GeV/c  hadrons, the results from Fig. 3 show that protons are not suppressed at any beam energy and pions go from being suppressed at higher beam energies to being enhanced at lower beam energies with a transition near $\sqrt{s_{_{ \mathrm{NN}}}}$ = 27 GeV.

\section{Summary} 
Charged hadron $R_{ \mathrm{CP}}$ has been measured at mid-rapidity for a range of beam energies in order to determine at what beam energy the suppression of high $p_{ \mathrm{T}}$ charged hadrons, a QGP signature, turns off.  Suppression is seen to turn off near 39 GeV for unidentified charged hadrons, but due to unquantified sources of enhancement it is currently unclear where medium-induced jet quenching turns off.   Identified $R_{ \mathrm{CP}}$ is qualitatively similar and promotes pions as a probe that is less effected by sources of enhancement.  For pions, suppression at high $p_{ \mathrm{T}}$ is seen to turn off near 27 GeV. A comparison to the HIJING event generator with jet quenching turned off did not reveal an energy at which the data and the model agree quantitatively, which precludes measuring the energy at which data and simulation deviate.  This motivates the exploration of additional tunes of HIJING and other models in order to disentangle the competing effects which lead to suppression or enhancement.

\end{document}